\documentclass[fleqn,10pt]{wlscirep}
\usepackage[utf8]{inputenc}
\usepackage[T1]{fontenc}
\title{Enhanced photon emission from a double-layer target at moderate laser 
intensities}

\author[1,2*]{Martin Jirka}
\author[1,2]{Ondrej Klimo}
\author[1,3]{Yan-Jun Gu}
\author[1,4]{Stefan Weber}
\affil[1]{Institute of Physics of the CAS, ELI-Beamlines Project, Na Slovance
	2, Prague, 182 21, Czech Republic}
\affil[2]{Faculty of Nuclear Sciences and Physical Engineering, Czech
	Technical University in Prague, Brehova 7, Prague, 115 19, Czech Republic}
\affil[3]{Institute of Plasma Physics of the CAS, Za Slovankou 1782/3, Prague, 
182 00, Czech Republic}
\affil[4]{School of Science, Xi’an Jiaotong University, Xi’an, 710049, China}

\affil[*]{martin.jirka@eli-beams.eu}

\begin{abstract}
	In this paper we study photon emission in the interaction of the laser 
	beam with an under-dense target and the attached reflecting plasma mirror.
	Photons are emitted due to the inverse Compton scattering when accelerated electrons interact with a reflected part of the laser pulse.
	The enhancement of photon generation in this configuration lies in using 
	the laser pulse with a steep rising edge.
	Such a laser pulse can be obtained by the preceding interaction of the incoming laser pulse with a thin solid-density foil.
	Using numerical simulations we study the origin of such a laser pulse and 
	its effect on photon emission.
	As a result, accelerated electrons can interact directly with the most intense part of the laser pulse that enhances photon emission. 
	This approach increases the number of created photons and improves photon beam divergence.
\end{abstract}
\begin{document}

\flushbottom
\maketitle
%
%
\thispagestyle{empty}

\section*{Introduction}
Using today's laser systems, electrons can be accelerated up to 8~GeV in 20~cm long capillary discharge waveguide \cite{Gonsalves2019}.
When a bunch of accelerated electrons collides head-on with an intense laser pulse, these electrons will emit high-energy photons due to the inverse Compton scattering \cite{Cole2018,Poder2018}.
The goal is therefore the realization of the all-optical compact source of high-energy $ \gamma $-ray beam \cite{Shvets2011,TaPhuoc2012}.
However, the efficiency of such a source depends on the properties of the 
electron and laser beams and on precise alignment of their interaction.
Probability of photon emission is characterized by the parameter $ 
\chi_{e}=\gamma/E_{\mathrm{S}}\sqrt{(\mathbf{E}+\mathbf{v}\times\mathbf{B})^{2}-(\mathbf{v}\cdot\mathbf{E}/c)^{2}}
 $, where $ \gamma $ is the relativistic factor of the emitting particle 
(electron), $ \mathbf{E} $ and $ \mathbf{B} $ are the electric and magnetic 
fields, $ \mathbf{v} $ is the particle velocity and $ c $ is the speed of light 
in SI units \cite{Ritus1985}.
In previous equation $ E_{\mathrm{S}} $ is the Sauter (Schwinger) limit field $ E_{\mathrm{S}}=m_{e}^{2}c^{3}/(e\hbar)\approx1.33\times10^{18}~\mathrm{V/m} $, $ m_{e} $ is the electron rest mass, $ e $ is the elementary (positive) charge and $ \hbar $ is the reduced Planck constant \cite{Sauter1931,Schwinger1951}.
This parameter is maximized, when the electron is colliding head-on with the laser pulse.
In such a case, the value of $ \chi_{e} $ can be approximated as
$ \chi_{e}\approx2\gamma E_{0}/E_{\mathrm{S}} $, where $ E_{0} $ is the amplitude of the laser field \cite{Bulanov2013}.
Then photon emission probability is only controlled by the energy of the 
incoming electron and the amplitude of the laser field.

Electrons in plasma can be accelerated by Laser Wake-Field Acceleration (LWFA) or Direct Laser Acceleration (DLA) mechanisms \cite{Tajima1979,Pukhov1999,Gahn1999}. 
The latter becomes more important in the case of plasma densities higher than $ 
10^{20}~\mathrm{cm^{-3}} $ and intensities going beyond today's world record 
($> 10^{22}~\mathrm{W/cm^{2}} $) 
\cite{Pukhov2002,Shaw2014,Bahk2004,Yanovsky2008,Pirozhkov2017}.
To achieve such a high intensity, the laser pulse has to be tightly focused that will result in rapid diffraction of the laser field.
Thus, the higher plasma density is required to compensate for diffraction in 
this case.

Nevertheless, head-on collision remains an issue from the experimental point of view due to the spatio-temporal alignment of the interaction \cite{Cole2018}.
This can be overcome by employing a plasma mirror.
As the laser pulse impinges on the over-dense plasma mirror, it is reflected and thus 
previously accelerated electrons can interact with a counter-propagating laser 
field that leads to efficient photon emission \cite{TaPhuoc2012,Yu2016}.
This double-layer interaction setup can be further optimized by tuning the 
target properties (density, thickness) with respect to the laser intensity and 
focal spot radius to create the highest number of high-energy photons 
\cite{Gu2018Brilliant,Gong2018,Gu2018Intense,Huang2018,Huang2019,Liu2019,Long2019,Ong2019,Gu2019}.

\begin{figure}
	\centering
	\includegraphics[width=0.75\linewidth]{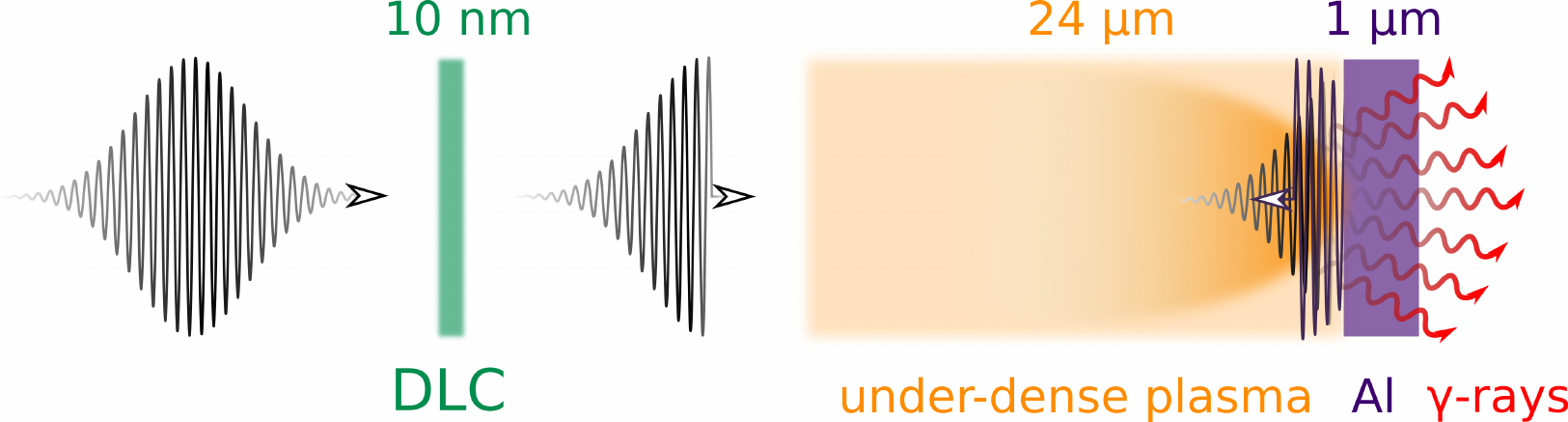}
	\caption{Interaction setup: the laser pulse gains a steep front edge after 
	passing through the Diamond-Like-Carbon (DLC) layer (green). In the next 
	stage, it accelerates electrons in the under-dense target and collides with 
	them as it is reflected from Al plasma mirror. As a result of this 
	interaction, $ \gamma $-ray photons are emitted.}
	\label{fig:Fig_01}
\end{figure}

In this paper, we study photon emission in such an interaction scheme when various temporal profiles of the incoming laser pulse are assumed.
For efficient photon production in a laser-electron collision is crucial for the 
electron to get in the highest intensity region.
As the electron enters the laser filed, it starts losing energy and thus can be expelled by the ponderomotive force before reaching the laser field amplitude.
This effect that acts against efficient photon emission can be overcome by employing a tailored temporal profile 
of the laser pulse.
The laser pulse with a steep front edge ensures that accelerated electrons will 
interact directly with the most intense part of the laser pulse and that 
consequently enhances photon emission, see Fig.~\ref{fig:Fig_01}.
Using numerical simulations we study the origin of such a laser pulse and its effect on photon emission.  

\section*{Results}
%

To analyze photon emission in this interaction setup, we have performed 2D 
Particle-In-Cell (PIC) simulations in the code EPOCH 
\cite{Ridgers2014Modelling,Arber2015}.
At first, we considered the interaction of the laser pulse with $ 
24~\mathrm{\mu m} $-thick under-dense 
target containing 
electrons and protons of a density $ 0.1n_{\mathrm{c}} $, where 
$ 
n_{\mathrm{c}}=\omega_{0}^{2}m_{e}\epsilon_{0}/e^{2} $ is the critical electron 
density and $ 
\epsilon_{0} $ is the vacuum permittivity.
At the rear side of the under-dense target, $ 1~\mathrm{\mu m} $-thick $ 
\mathrm{Al}^{11+} $ foil of the electron density $ 385n_{\mathrm{c}} $ is 
attached.
This part of the double-layer target serves as a reflecting mirror for the 
laser pulse. 

The incoming laser pulse has a wavelength of $ 805~\mathrm{nm}$ and 
Full-Width-At-Half-Maximum duration of $ \tau=30~\mathrm{fs}$.
The peak intensity $ I_{0}=5\times10^{21}~\mathrm{W/cm^{2}} $ of the focused 
laser beam corresponds to the normalized laser amplitude $ 
a_{0}= eE_{0}/(m_{e}\omega_{0}c)=45 $ where $ \omega_{0} $ is the laser angular 
frequency.
These laser parameters are well within the capabilities of today's laser 
systems such as J-Karen-P 
\cite{Pirozhkov2017}.
We compared the interaction in which the laser pulse had either the 
Gaussian (Setup I) or perfectly tailored (Setup II) temporal duration.
The latter case was modelled by cutting the front edge of the laser 
pulse so 
that the 
electric field was equal to zero up to one-quarter of 
the laser period 
before 
the 
peak amplitude. 
Such a beam therefore delivers by almost 50\% less energy onto the target compared to Setup I.
Depending on the temporal profile of the laser pulse, the acceleration of 
electrons and consequently photon emission differ.

\begin{figure}[ht!]
	\centering
	\includegraphics[width=1.0\linewidth]{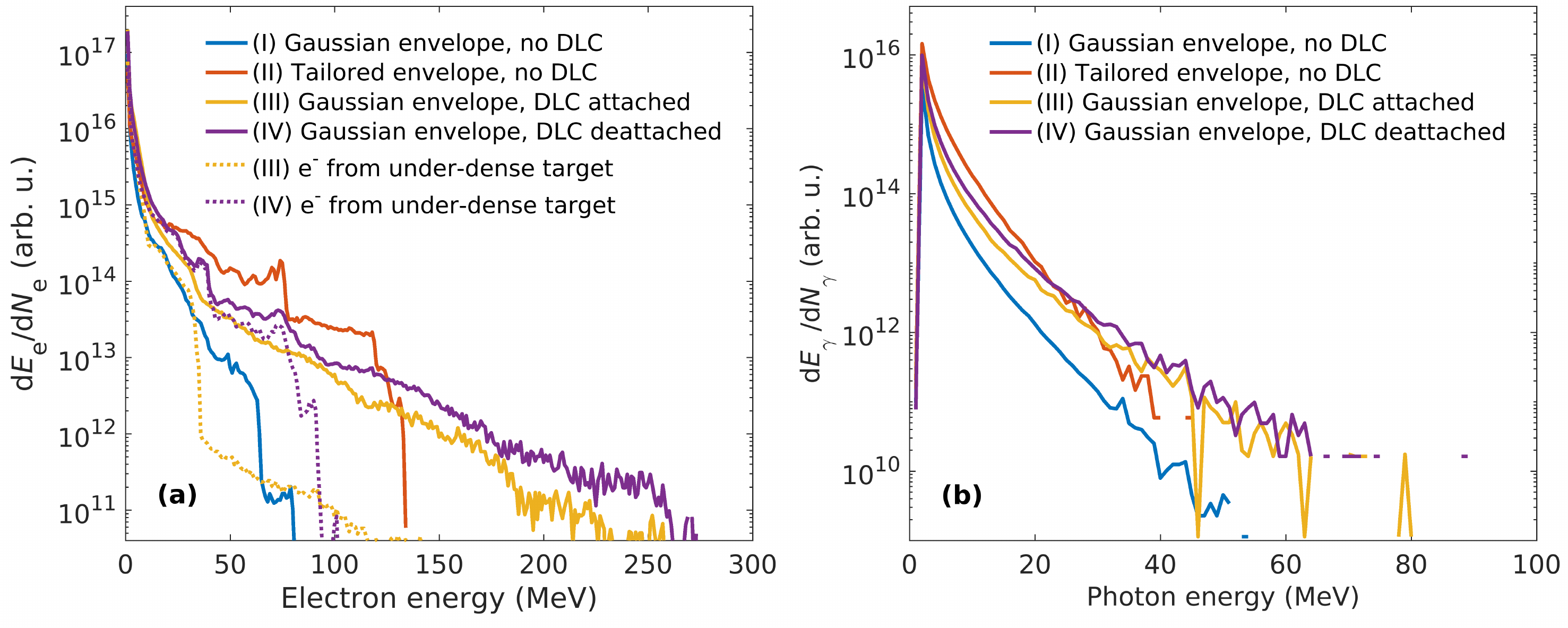}
	\caption{Energy distribution of (a) electrons at the time when the laser 
			beam reaches the end of the under-dense target and of (b) photons 
			at the end of the interaction.
			The laser beam has either (I) the Gaussian 
			or (II) tailored temporal profile; or the laser beam with the 
			Gaussian temporal 
			profile is assumed while the DLC layer is (III) attached or (IV) 
			detached from 
			the under-dense target. Dotted lines represent only electrons from 
			the under-dense target in corresponding runs.
}
	\label{fig:Fig_02}
\end{figure}

In Setup I, when the laser pulse has the Gaussian temporal envelope, the 
propagation 
distance of $ 24~\mathrm{\mu m} $ in under-dense plasma is not long enough for a 
wakefield to become fully developed.
By contrast, employing a tailored laser beam leads to considerable enhancement 
of electron acceleration.
This is represented by Setup II.
In such a case, the laser beam has both a shorter duration and a steeper rise 
of the front edge.
Therefore, the bubble behind the laser pulse develops more rapidly 
compared to the previous case as the laser pulse propagates through under-dense plasma.
That enables more efficient acceleration of electrons via LWFA mechanism and 
improves the injection of electrons into the laser pulse structure for DLA 
mechanism.
In Fig.~\ref{fig:Fig_02}a we show electron energy spectra at the time when 
the laser pulse reaches the end of the under-dense target.
By comparing lines I and II it is clear that much more electrons with higher 
energies are produced when the laser beam has a tailored temporal profile.

Photon emission can only be enhanced when these high-energy electrons 
collide with the laser field reflected from the attached aluminium foil.
In Fig.~\ref{fig:Fig_02}b we present the energy spectra distribution of 
generated photons during the interaction.
The cut-off energy of generated photons in Setup II is below 50~MeV even though electrons can be accelerated up to 130~MeV in this case, 
see Fig.~\ref{fig:Fig_02}a.
However, the electrons having the highest energy are trapped in the front part of the tailored laser pulse, thus these DLA electrons can not collide 
with a sufficiently long part of the reflected laser pulse.
High-energy photons are more likely generated by DLA electrons locked in the 
rear part of the 
laser pulse as well as by LWFA electrons dragged behind the laser pulse.
The results for all presented Setups are summarized in Tab.~\ref{tab:Tab_01}: 
efficiency of photon emission, i.e. number of photons, their mean energy and 
efficiency of laser energy conversion to photons.
Employing the laser pulse with a steep front edge results in 
generating 5x more photons.
The mean photon energy in such a case is about 40\% higher and conversion 
efficiency is increased by a factor of six compared to Setup I.
\begin{table}[h!] 
	\centering
	\caption{The number of photons $ N_{\gamma} $, their mean energy $ 
	\left\langle 
		E_{\gamma} 
		\right\rangle  $ and conversion efficiency $ \eta_{\gamma} $ of laser 
		energy to 
		photons for simulation Setups I--IV. Energy is normalized to the 
		energy 
		of the Gaussian laser pulse.}
	\begin{tabular}{c|cccc}
		& I & II & III &  IV \\ 
		\hline 
		$ N_{\gamma}~(1/\mathrm{m}) $  & $ 5.2\times10^{15} $ & $ 
		2.5\times10^{16} $  
		& $ 
		1.2\times10^{16} $ & 
		$ 
		1.5\times10^{16} $
		\\ 
		$ \left\langle E_{\gamma} \right\rangle (\mathrm{MeV}) $ & 1.52 & 2.12 
		& 1.66 & 
		1.89 \\ 
		$ \eta_{\gamma}~(\%)  $ & 0.035 & 0.23 & 0.088 & 0.12
		\\ 
	\end{tabular} 
	\label{tab:Tab_01}
\end{table}

The laser beam with a steep front edge can be realized by the interaction of the laser 
pulse with a thin 
solid-density foil \cite{Vshivkov1998,Palaniyappan2012,Wei2017,Qu2017}.
Since the foil is over-dense for the incoming laser pulse, the front part of the 
laser pulse is reflected.
As the peak of the laser pulse impinges upon the foil surface, the 
relativistic 
mass of electrons suddenly increases that, in turn, causes the foil to become 
relativistically transparent for the rest of the laser pulse.
Therefore, the laser pulse gains a steep front edge after passing through the foil.

\begin{figure}[ht!]
	\centering
	\includegraphics[width=0.5\linewidth]{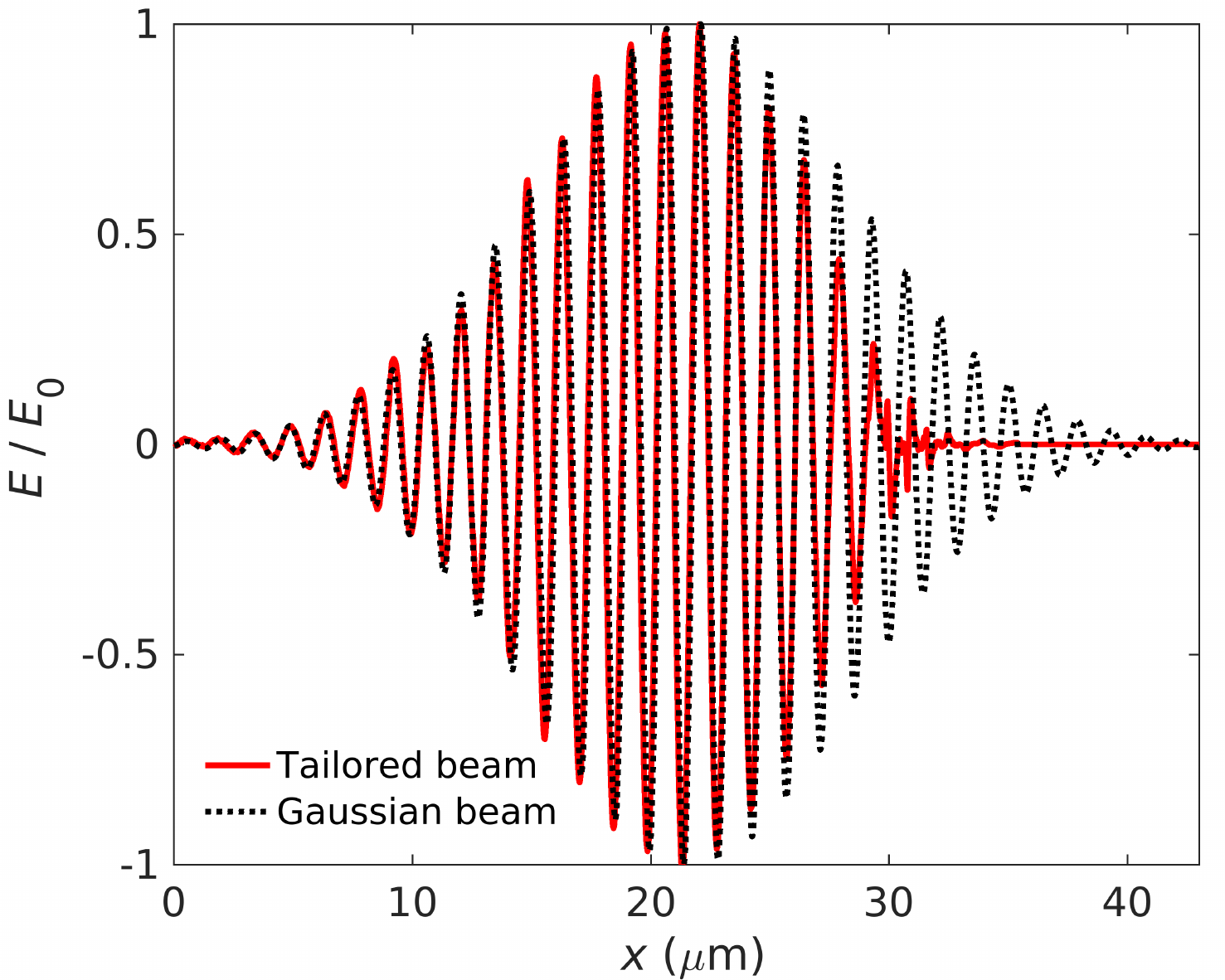}
	\caption{The result from 2D PIC simulation: $ E_{y} $ component of the 
	laser 
	pulse 
	before (black) and after (red) passing through the DLC layer.}
	\label{fig:Fig_03}
\end{figure}

\begin{figure}[ht!]
	\centering
	\includegraphics[width=0.6\linewidth]{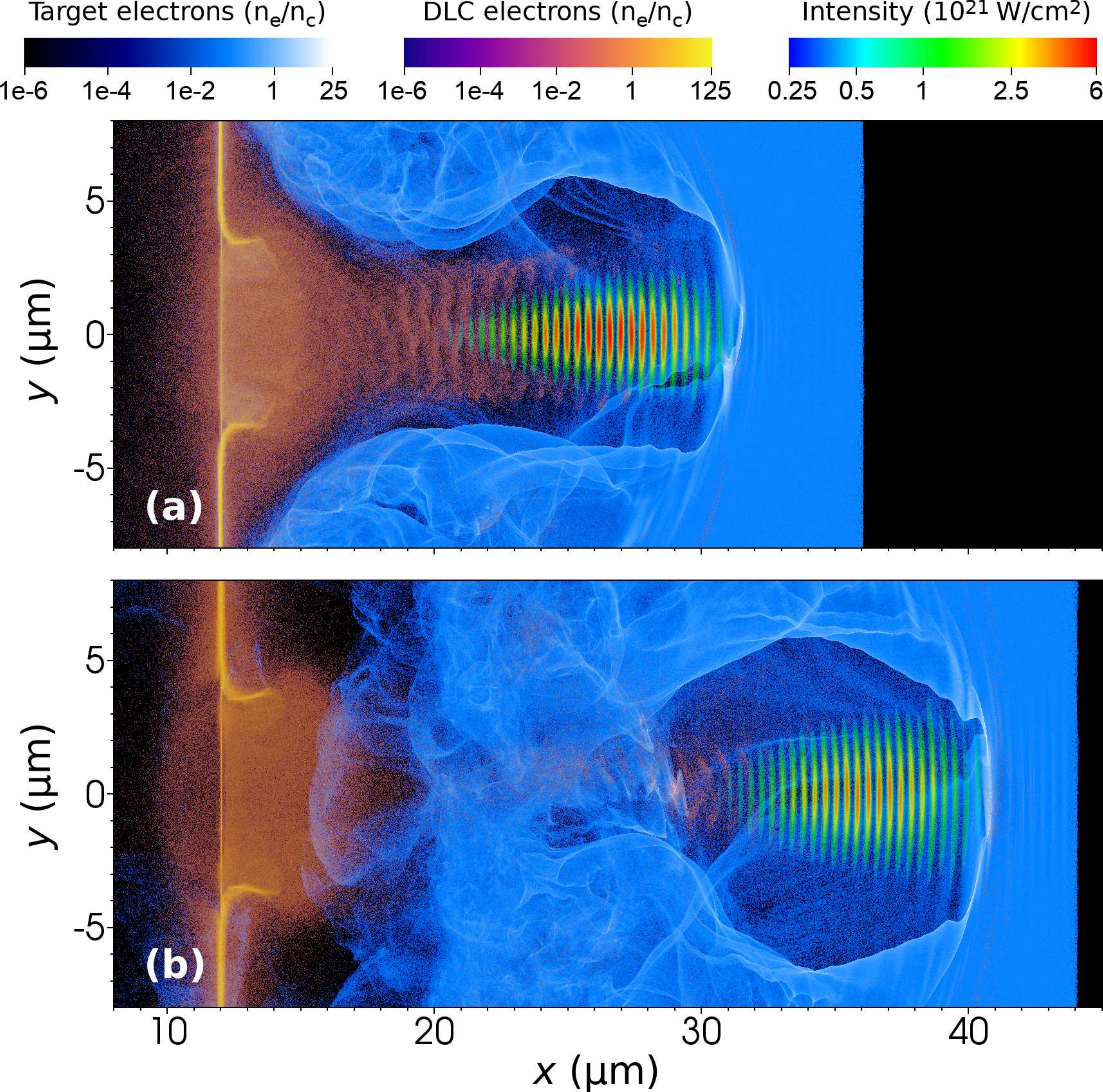}
	\caption{The density of DLC (orange) and target (blue) electrons in 
	simulation Setup 
		III and IV in which the DLC 
		layer is either (a) attached or (b) detached from the under-dense 
		target, respectively. The 
		incoming laser pulse has the Gaussian temporal
		profile.}
	\label{fig:Fig_04}
\end{figure}

To model this interaction, we have performed another simulation, Setup III, in which a $ 
10~\mathrm{nm} $-thin Diamond-Like-Carbon (DLC) foil is attached at the front 
side of the target \cite{Henig2009,Ma2011}.
The fully ionized DLC foil has the electron density $ 384n_{\mathrm{c}} $.
As the laser pulse initially having the Gaussian temporal profile passes through the DLC foil, it gets a 
steep front edge, as shown in Fig.~\ref{fig:Fig_03}.
By cutting the front part of the laser pulse, it loses about $ 15\% $ of its 
initial energy.
Then the tailored laser pulse interacts with the double-layer target.
In this case, the number and mean energy of created photons are much lower than 
in 
Setup II, see Tab.~\ref{tab:Tab_01}, as the presented DLC foil is not dense 
enough 
to create 
ideally tailored laser beam.
However, the number and mean energy of created photons are still higher 
compared to Setup I.
Nevertheless, the dynamics of DLC electrons negatively affects the 
acceleration of electrons originating in under-dense plasma.
These electrons are immediately expelled by the laser pulse while the DLC ones are attracted by protons to compensate for the charge separation field created behind the laser pulse.
For this reason, the electrons which are expelled sideways by the 
ponderomotive force can not form the bubble and thus are not trapped at 
the back of this structure, see Fig.~\ref{fig:Fig_04}a.
Thus, the acceleration of electrons in under-dense plasma is efficiently 
reduced.
It agrees with the electron spectrum presented in Fig.~\ref{fig:Fig_02}a 
showing 
that in Setup III electrons belonging to the under-dense target (dotted line) 
have 
much lower cut-off energy than the DLC ones.
In Setup III, the cut-off energy for photons is 80~MeV.
As shown in Fig.~\ref{fig:Fig_02}a, there are much more DLC electrons with 
energy higher than 50~MeV.
Therefore, mainly the DLC electrons located in the rear part of the laser pulse are the ones responsible for generation of high-energy photons.

As the main disadvantage of the previous interaction setup is that the laser 
wake-field structure can not fully develop in the under-dense target we 
propose another configuration, 
Setup IV,
in 
which the DLC layer is initially detached from the double-layer target by a $ 
8~\mathrm{\mu m} $ vacuum gap.
Due to the vacuum gap between the DLC layer and the under-dense target, the DLC 
electrons do not 
prevent forming of the bubble in plasma, as shown in 
Fig.~\ref{fig:Fig_04}b.
It leads to a more efficient LWFA of electrons originating their motion in 
the under-dense target.
These electrons are trapped behind the laser pulse and thus they have the 
favourable 
position for emitting photons when they interact with the reflected laser pulse.
As a result, more photons are emitted compared to Setup III as shown in 
Fig.~\ref{fig:Fig_02}b.
However, the cut-off in the photon energy spectrum is still about 80~MeV 
despite 
the improvement in the electron energy spectrum cut-off.
This is due to the fact that the most energetic electrons are the DLC ones locked in the front part of the laser pulse which do not have a chance to significantly 
contribute to photon emission.
Nevertheless, employing the detached DLC layer allows creating the highest 
number of high-energy 
photons in comparison with all the Setups presented above, see 
Fig.~\ref{fig:Fig_02}b.
The theory predicts photons with typical energy around 2~MeV for Setup IV  
\cite{Bell2008}.
This agrees with our results from PIC simulations presented in 
Tab.~\ref{tab:Tab_01}.
The conversion of the laser energy to photons is increased by a factor of 3.4 compared 
to Setup I.
Optimizing the distance between the DLC layer and the target with respect to the 
target density can further enhance conversion efficiency.
The length of a vacuum gap allows the DLC electrons to expand and thus reduce 
their number which enters the under-dense target.
For example, in our case the optimal length of the vacuum gap is 
about $ 4~\mathrm{\mu m} $ according to the results of PIC simulations.
The conversion efficiency of laser energy to photons in such a case is four 
times higher compared to Setup I.

\begin{figure}
	\centering
	\includegraphics[width=0.7\linewidth]{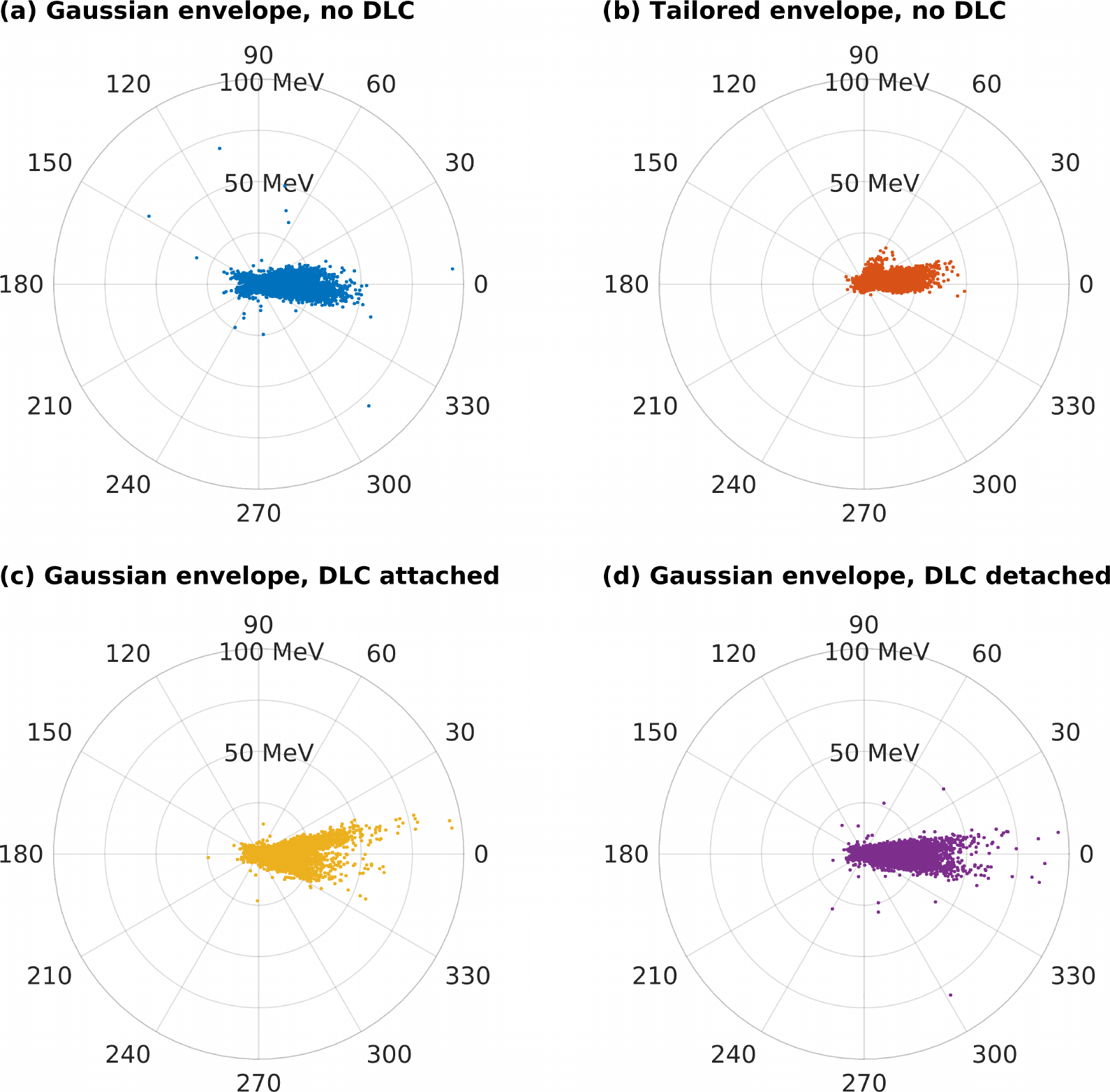}
	\caption{The angular energy distribution of photons for Setups I--IV at 
	the end 
		of the simulation. The laser beam has 
					either (a) the Gaussian 
					or (b) tailored temporal profile; or the laser beam with 
					the 
					Gaussian temporal 
					profile is assumed while the DLC layer is (c) attached or 
					(d) 
					detached from 
					the under-dense target.
	}
	\label{fig:Fig_05}
\end{figure}

Moreover, the angular characteristics of the 
emitted photon beam are also 
improved,  see 
Fig.~\ref{fig:Fig_05}.
When the DLC layer is employed and detached from the target (Setup IV), the 
LWFA mechanism 
can develop that results in a narrower 
angular distribution of emitted photons compared to Setup III, compare 
Figs.~\ref{fig:Fig_05}(c,d).

\section*{Discussion}
Up to this point, we have assumed the fixed target density while the distance 
from the DLC layer was varied.
Increasing the plasma density leads to a creation of a higher number of photons 
by 
electrons from the under-dense target and thus more efficient laser energy 
conversion provided that the 
relativistic critical density $ \gamma n_{\mathrm{c}} $ is not reached.
However, if the plasma density is too high then the laser pulse can be rapidly 
depleted.
On the other hand, if an intense laser pulse propagates in  
near-critical-density plasma for a sufficiently long distance, it may undergo 
the effect of relativistic self-focusing that increases the laser intensity and 
reduces diffraction \cite{Sun1987,Mourou2006}. 
That, in turn, can lead to emission of photons with higher energy.
The optimal propagation distance with respect to a given plasma density is 
therefore limited by these two effects 
\cite{Bin2015,Huang2016Characteristics,Huang2019}.

To assess the role of a higher plasma density, we have performed additional 
simulations, in which the target density 
has been increased by a factor of 10 from $ 0.1n_{\mathrm{c}} $ to $ 1n_{\mathrm{c}} $.
Targets of such a density have been already demonstrated, e.g. 
Refs.~\cite{Qiao2017,Ma2019}.
Due to self-focusing, the laser pulse gains a smaller transverse profile and a higher peak intensity.
In our case, the peak intensity of the laser field in $ 1n_{\mathrm{c}} $ 
target is by 25\% higher than in $ 0.1n_{\mathrm{c}} $ one.
Moreover, the DLA is more efficient at such a plasma density as it enables to 
accelerate a higher number of electrons.
As a result, the laser energy conversion to photons is higher by a factor of 15 
compared to Setup I, i.e. when the DLC layer is not considered.
Employing the DLC layer is still feasible for such a dense target as it 
enhances photon production by a factor of 1.3.
This confirms the applicability of our setup even for near-critical-density 
plasma targets.

The efficiency of laser energy conversion to photons in Setup IV is approximately the same as in the case when 20~pC LWFA electron bunch having energy 0.5~GeV collides with the laser pulse of the same properties as described above.
In such a case we obtain $ \eta_{\gamma}\approx0.10\% $ according to Ref. 
\cite{Vranic2014}.
Although it is possible to achieve higher electron energies using LWFA compared 
to our setup, it might be complicated to reflect and focus the driving laser 
pulse to initiate photon emission \cite{Feng2017,Ji2019}.
The presented setup is therefore more robust as it encompasses both the 
acceleration and photon-emission stages while the latter does not rely on a 
focusing mirror.

It has been shown that photon emission in the interaction of a laser pulse with the under-dense target and reflecting plasma mirror can be enhanced by employing a laser pulse with a steep front edge.
Such a beam can be provided by the preceding interaction of the laser pulse with a thin solid-density foil.
The shaped laser pulse then propagates through the vacuum into the under-dense  
target in which electrons are accelerated via LWFA and DLA mechanisms.
The vacuum gap between the foil and the target ensures that electrons dragged 
from this foil will not counteract the acceleration of electrons in the 
under-dense target.  
The accelerated electrons then interact with the most intense part of the laser pulse reflected from the plasma mirror.
Therefore, employing the solid-density foil will result in a more efficient conversion of the laser energy to photons.
For the parameters described above we obtained three times higher 
conversion 
efficiency and a narrower angular distribution of emitted photons compared to 
the interaction without the thin solid-density foil.
This can be further improved by adjusting the density and thickness of the 
foil to provide the optimal temporal profile of the laser pulse.
As the laser pulse loses its energy in the under-dense target very slowly, the 
length of the electron acceleration stage could be optimized with respect to 
the laser intensity to get the highest number of accelerated electrons.

\section*{Methods}
\subsection*{Numerical Modelling}
To analyze the presented laser-plasma interaction we used the PIC code 
EPOCH in which photon emission is considered as a step-like quantum 
process \cite{Arber2015}.
For details about the implementation of photon emission into this code the 
reader 
is referred to Ref. 
\cite{Ridgers2014Modelling}.

In 2D simulations of Setups I--IV, the box was spanning from $ 
0 $ to $ 50~\mathrm{\mu m} $ in the $ x $-direction and from $ -15~\mathrm{\mu 
m} $ to $ 15~\mathrm{\mu m} $ in the $ y $-direction.
Such a simulation domain was resolved with $ 22,320\times13,392 $ cells.
This is sufficient as for a density of $  385n_{\mathrm{c}} $ the plasma skin 
depth is about 6.5~nm.
The spatial resolution remained unchanged for all other performed 2D 
simulations  
(e.g. parameter scan for the optimal length of a vacuum gap), while the 
size 
of the simulation box was enlarged.
The laser pulse enters the box at a boundary $ x=0~\mathrm{\mu m} $.
The DLC layer of thickness 10~nm was located at $ x=11.99~\mathrm{\mu m} $ 
(Setups III and IV) while
the under-dense 
target was spanning from $ 12~\mathrm{\mu m} $ to $ 36~\mathrm{\mu m} $ (Setups 
I--III) or from $ 20~\mathrm{\mu m} $ to $ 44~\mathrm{\mu m} $ (Setup IV).
At the rear side of the under-dense target a $ 1~\mathrm{\mu m} $-thick $ 
\mathrm{Al}^{11+} $ foil was attached.
The laser pulse having the Gaussian temporal envelope propagates in the 
positive $ x 
$-direction while being polarized along 
the $ y $-axis.
It is focused to a focal spot of radius $ w_{0}=1.5~\mathrm{\mu m} $ located at 
$ x=12~\mathrm{\mu m} $ in the simulation box.

\section*{Data Availability Statement}
The datasets generated and analyzed during the current study are available from 
the corresponding author on reasonable request.

\bibliography{main}{}

\section*{Acknowledgements}

This work is supported by the project High Field Initiative 
(CZ.02.1.01/0.0/0.0/15\_003/0000449) from European Regional Development Fund.
The support of Czech Science Foundation project No. 18-09560S is acknowledged.
The results of the  Project LQ1606 were obtained with the financial support of 
the Ministry of Education, Youth and Sports as part of targeted support from 
the National Programme of Sustainability  II.
The EPOCH code used in this research was developed under UK EPSRC grants 
EP/G054950/1, EP/G056803/1, EP/G055165/1 and EP/M022463/1.
The simulations were performed at the cluster ECLIPSE at ELI Beamlines.

\section*{Author contributions statement}

M.J. carried out the simulations, analyzed the results and wrote the bulk of 
the manuscript. O.K., Y-J.G. and S.W. provided supervision
and theoretical support for the interpretation of the results. All authors 
contributed to the preparation of the manuscript.

\section*{Additional information}

\textbf{Competing interests}: The authors declare that they have no competing 
interests.

\end{document}